\documentclass[aps,prb,twocolumn,groupedaddress]{revtex4}
\usepackage[dvipdfm]{graphicx}
\usepackage{bm}
\usepackage{amsmath}

\begin{document}


\title{
Two Phase Collective Modes in Josephson Vortex Lattice in
Intrinsic Josephson Junction Bi$_2$Sr$_2$CaCu$_2$O$_{8+\delta}$ }



\author{I. Kakeya}
\email[]{kakeya@ims.tsukuba.ac.jp}
\affiliation{Institute of Materials Science, University of
Tsukuba, Tsukuba, Ibaraki 305-8573 Japan}
\author{T. Wada}
\affiliation{Institute of Materials Science, University of
Tsukuba, Tsukuba, Ibaraki 305-8573 Japan}
\author{R. Nakamura}
\affiliation{Institute of Materials Science, University of
Tsukuba, Tsukuba, Ibaraki 305-8573 Japan}
\author{K. Kadowaki}
\affiliation{Institute of Materials Science, University of
Tsukuba, Tsukuba, Ibaraki 305-8573 Japan}


\date{May 31, 2005}
\begin{abstract}
Josephson plasma excitations in the high $T_c$ superconductor
Bi$_2$Sr$_2$CaCu$_2$O$_{8+\delta}$ have been investigated in a
wide microwave frequency region (9.8 -- 75 GHz), in particular, in
magnetic field applied parallel to the $ab$ plane of the single
crystal. In sharp contrast to the case for magnetic fields
parallel to the $c$ axis or tilted from the $ab$ plane, it was
found that there are two kinds of resonance modes, which are split
in energy and possess two distinctly different magnetic field
dependences. One always lies higher in energy than the other and
has a shallow minimum at about 0.8 kOe, then increases linearly
with magnetic field. On the other hand, another mode begins to
appear only in a magnetic field (from a few kOe and higher) and
has a weakly decreasing tendency with increasing magnetic field.
By comparing with a recent theoretical model the higher energy
mode can naturally be attributed to the Josephson plasma resonance
mode propagating along the primitive reciprocal lattice vector of
the Josephson vortex lattice, whereas the lower frequency mode is
assigned to the novel phase collective mode of the Josephson
vortex lattice, which has never been observed before.
\end{abstract}


\maketitle


\section{Introduction\label{}}
The phase collective excitation in Josephson junctions known as
Josephson plasma has been studied since the 1960's, triggered by a
theoretical prediction of Anderson.~\cite{And64} However, there
are only a few experimental reports on Josephson plasma resonance
in single junctions made by conventional
superconductors~\cite{Dah68,Ped72} because there is a strong
damping effect of the quasiparticles due to small superconducting
energy gap in the single junctions. This situation was improved
drastically in the case of high $T_c$ superconductors because they
are comprised of well-established weakly coupled Josephson
junctions along the crystallographic $c$ axis (intrinsic Josephson
junction), whose plasma frequency  along the $c$ axis is much
lower than the superconducting gap. In particular, in the case of
Bi$_2$Sr$_2$CaCu$_2$O$_{8+\delta}$ (Bi2212),~\cite{Kle92,Kle94}
the superconducting CuO$_2$ layers and the insulating or
semiconducting Bi$_2$O$_2$ layers are regularly stacked in an
atomic level in the unit cell forming a weakly coupled Josephson
junctions. Based on this, Tachiki and co-workers predicted the
existence of the Josephson plasma excitation in high $T_c$
superconductors.~\cite{Tac94,Bul95} Subsequently, two theoretical
models have been developed : one by Koyama and
Machida,~\cite{Koy96,Mac99} the other by Pedersen and
Sakai.~\cite{Ped98,Sak99} Quite recently, a unified theory has
been reported.~\cite{Mac04}

The Josephson plasma resonance in Bi2212 is observed as a sharp
and strong microwave absorption in a finite magnetic field
parallel to the $c$ axis.~\cite{Mat95} Especially, the
longitudinal Josephson plasma mode gives an extremely sharp
resonance because of its small dispersion and the high
quasiparticle damping rate of Bi2212~\cite{PRB97,Oha99}. The clear
observation of the longitudinal plasma resonance has been a strong
advantage of Bi2212 for investigation of the Josephson plasma
phenomena in superconductors compared with many other materials
such as La$_{2-x}$Sr$_x$CuO$_4$,~\cite{Tam92}
Bi$_2$(Sr,La)$_2$CuO$_{6+\delta}$(Bi2201),~\cite{Sak96}
YBa$_2$Cu$_3$O$_{7-\delta}$(YBCO),~\cite{Taj97} and BEDT-TTF
salts,~\cite{Shi97} in which the Josephson plasma resonance (edge)
is less clear.

According to the theory of the Josephson plasma resonance, the
resonance frequency $\omega_p(H,T)$ can be written
as~\cite{Bul95,Bul96a}
\begin{equation}
\omega_p^2(H,T)=\omega_p^2(T) \langle \cos \varphi_{l,l+1}(H,T)
\rangle,\label{JPR}
\end{equation} where
$\omega_p(T)=c/\sqrt{\epsilon}\lambda_c(T)$ is the Josephson
plasma frequency in the absence of a magnetic field. $\epsilon$
and $\lambda_c(T)$ stand for the dielectric constant and the
temperature-dependent $c$ axis penetration depth, respectively.
$\varphi_{l,l+1}(H,T)$ is the gauge-invariant phase difference
between the ($l$)th and ($l+1$)th layers and $\langle \cdots
\rangle$ denotes the spatial and time averages.

In the case of a magnetic field parallel to the $c$ axis
(perpendicular to the superconducting CuO$_2$ layers), it is well
known that pancake vortices are generated above $H_{c1}$, the
lower critical field. For the Josephson plasma resonance, this
situation is taken into account by considering the uniform
reduction of the Josephson current along the $c$-axis, resulting
in the decrease of $\langle \cos \varphi_{l,l+1}(H,T) \rangle$,---
{\it i.e.}, the reduction of $\omega_p(H,T)$ in Eq. (\ref{JPR}).
Because of this fact, the Josephson plasma resonance can be used
as a sensitive method to evaluate $\langle \cos
\varphi_{l,l+1}(H,T)
\rangle$.~\cite{Kos96,Shi99a,plasma2ka,Kos00b}

This treatment cannot be applied to the case where the external
magnetic field is parallel to the $ab$ plane. In this case, the
Josephson vortices are introduced in-between the CuO$_2$ double
layers and contribute the Josephson plasma resonance as an
essential ingredient of the phenomenon. In contrast to the case
for the perpendicular field, the plasma frequency is hardly
suppressed because the fluctuations of pancake vortices are too
small to dominate the interlayer coupling. This is clearly
observed in the angular dependence of the Josephson plasma
resonance near the $ab$ plane. A sharp dip in the Josephson plasma
resonance field is clearly observed within an angle of $\pm$ 2
degrees from the $ab$ plane.~\cite{Plasma97a} From this sharp
change of the Josephson plasma resonance behavior, it is in turn
expected that the dynamical motion of Josephson vortices such as
Josephson vortex lattice collective modes has been thought to
account for the Josephson plasma resonance frequency. This opens
entirely new possibilities for the Josephson plasma resonance
phenomena in the case of parallel magnetic field configurations
and deserves more detailed consideration both experimentally and
theoretically. Since the Josephson vortex modulates the interlayer
Josephson tunnelling current in the length scale of
$\lambda_J\equiv\gamma s$ along the $ab$ plane, where $\gamma$ is
the anisotropy parameter and $s$ is the interlayer distance, the
collective plasma oscillation strongly couples with the collective
Josephson vortex motion such as Josephson vortex lattice modes.

This picture is initiated by the single junction model, which was
studied in the 1960s by Lebwohl and Stephen~\cite{Leb67} and by
Fetter and Stephen~\cite{Fet68} theoretically. They solved a one-
dimensional nonlinear wave equation and obtained two modes as
excitation spectra: one of them lies above the plasma gap
$\hbar\omega_p$ in zero field and approaches asymptotically the
linear relation with the propagation vector $\bm{k}$ (plasma
mode). Another mode obtained by Fetter and Stephen in a single
junction is a gapless vortex sliding mode. The plasma mode with an
array of Josephson vortices was first of all observed indirectly
as a so-called Eck resonance in the current--voltage ($I-V$)
characteristics of the Pb/PbO/Pb single junction, in which the
resonance was detected at a voltage proportional to the applied
parallel field $H_{\parallel}$.\cite{Eck64} Fiske has also shown
an anomalous step-like behavior in the $I-V$ characteristics in
Al, Sn, Pb, and Nb junctions~\cite{Fis64} and this phenomenon was
associated with the resonant electromagnetic modes of the
junction.

More recently, Josephson plasma resonance experiments have been
performed in the external magnetic field configuration being
closely parallel to the $ab$ plane in single-crystal
Bi2212.~\cite{Mat97a,Tsu97,Plasma97a} They observed anomalous
behaviors such as the angular dependence of the resonance field
associated with the sudden jump of it, which have partly been
explained by the theory of Bulaevskii {\it et al.}~\cite{Bul97} It
seems that the resonance mode indicating this anomalous behavior
can be described qualitatively by the single junction
model~\cite{Fet68}. However, the model in which conventional
(metallic) single junctions are connected in series cannot be
applied in intrinsic Josephson junctions especially in treating
Josephson plasma excitations because the charge conservation
inside a layer does not hold due to thinner CuO$_2$ double layers
(3 \AA) than the charge screening length ($\mu \sim 10$ \AA).
Therefore, we consider that the previous treatment of the
Josephson plasma resonance in a parallel magnetic field is
insufficient and exclude the rich physics inherent to the
intrinsic Josephson junctions.

In this paper, we report observation of two-phase collective modes
through systematic measurements of microwave absorption as
functions of magnetic fields parallel to the $ab$ plane,
temperature, and frequency in Bi2212 single crystals, and then the
origins of the two modes are argued. The organization is as
follows. Experimental results by using the microwave resonance
technique described in Sec. \ref{Experimental} are presented in
Sec. \ref{Results}. In this section, we first present the
experimental data at 25.5 GHz, where the phenomena are the most
interesting and the richest among all frequencies (Sec.
\ref{sec:25GHz}). Next, we show the experimental results at
different frequencies and describe the frequency dependence in
order to obtain a field dependence of the resonance modes (Sec.
\ref{sec:Freq}). Finally, universal features deduced from the
doping dependence are described (Sec. \ref{sec:Doping}). The
origins of the observed two modes are discussed in Sec.
\ref{Origin} with the help of recent theoretical
models.~\cite{Kos01,Mac01,Koy03}

\section{Experimental setup\label{Experimental}}
In order to perform Josephson plasma experiments in the Josephson
vortex state, it is necessary to have a wide range of microwave
frequencies beyond the zero-field plasma frequency, which is
expected above 100 GHz for optimally doped Bi2212. Since in our
experimental facility the microwave frequencies were limited up to
90 GHz, it was necessary to bring down the zero-field Josephson
plasma resonance in the range of 50 GHz. This was achieved by
shifting the doping level to the underdoped region by annealing
the pristine overdoped Bi2212 samples under reduced atmospheres.

We have measured microwave absorption in three Bi2212 single
crystals grown by the modified traveling solvent floating zone
(TSFZ) method. Two underdoped crystals (U1, U2) and one
optimallydoped crystal (OP) were used in the measurements for the
comparison. The superconducting transition temperatures $T_c$'s of
these crystals were determined by low-field magnetization
measurements with a superconducting quantum interference device
(SQUID) magnetometer as 70.2, 76.8, and 90.5 K with
superconducting transition widths of 1.0, 2.5, and 0.8 K for U1,
U2, and OP, respectively. The typical size of the crystals is $0.8
\times 0.8 \times 0.02$ mm$^3$.

Most of the microwave absorption data except for those in Sec.
\ref{sec:Doping} have been obtained in U1. Measurements were made
in a frequency range between 9.8 and 75 GHz by  both reflection-
and transmission-type bridge balance techniques using rectangular
cavity resonators with TE$_{102}$ mode. In order to excite the
Josephson plasma, the samples were placed inside the cavity in
such a way that the oscillating electric field of the microwaves
was exerted parallel to the $c$ axis over the $ab$ plane of the
crystal as shown in Fig. 1 of Ref.~\cite{PRB98}.
Frequency-stabilized microwaves were generated by the synthesized
signal generator (Hewlett Packard 83650B) or Gunn oscillators, and
the magnetic field was applied by a split-pair superconducting
magnet. The angle between field direction and CuO$_2$ plane of the
sample was adjusted by rotating the cavity resonator with respect
to the field direction by a precision rotator within an accuracy
of 0.001 degree.

The resonance data were obtained either by sweeping the magnetic
field at various fixed temperatures or by sweeping the temperature
at various fixed fields. The resonance occurs when $\omega_p(H,T)$
matches with the incident microwave frequency $\omega$ by varying
either the magnetic field or temperature. It is worth noting that
the field sweep (FS) measurements would give a non-equilibrium
Josephson vortex state because the strong hysteretic behavior of
the resonance was observed, especially at low temperatures while
the magnetic field was swept up and down. In order to avoid this
unnecessary complication, we employed measurements mostly by the
temperature sweep (TS) when low-temperature data were needed. The
temperature was either stabilized within 20 mK during the field
sweep or swept at a rate of 2 K/min. It is also worth noting that
since it is technically difficult to sweep microwave frequency as
far as the cavity resonator method is employed, we are compelled
to prepare a number of cavities (about 15 cavities) with different
resonance frequencies.

\section{Experimental Results\label{Results}}

\subsection{Simultaneous observation of two resonance modes
\label{sec:25GHz}}

\begin{figure}
\includegraphics[width=0.9\linewidth]{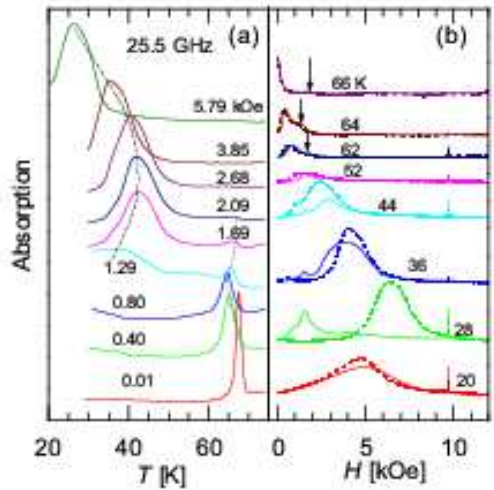}
\caption{
The resonance curves obtained at 25.5 GHz by sweeping temperature
(a) and by sweeping field (b). In (a), temperature was swept from
far above $T_c$ to the lowest temperature after changing magnetic
field at each measurement, and no hysteresis was found. In (b),
the solid and broken curves indicate data obtained in increasing
and decreasing fields. Hysteresis found in field sweep measurement
becomes more significant at lower temperatures and in lower
fields. Thick arrows in (b) denote shoulders attributed to the
weak resonance of HTM above the turnover. The sharp peak shown in
(b) at 10 kOe is due to DPPH as a field marker. }
\label{rawdata25}
\end{figure}

\begin{figure}
\includegraphics[width=\linewidth]{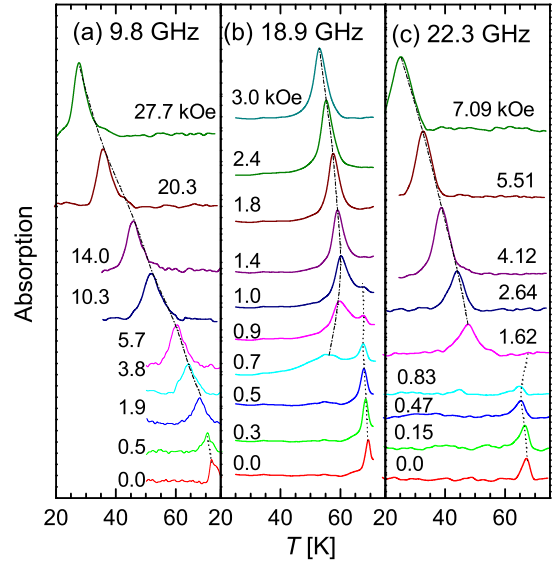}
\caption{
The resonance curves observed at 9.8 (a), 18.9 (b), and 22.3 (c)
GHz. The temperature is swept down from far beyond $T_c$ after
setting magnetic fields. At 9.8 GHz (a), the separation of LTM and
HTM cannot be identified although the line-shape changes
asymmetric to symmetric at a low field region below 0.5 kOe. At
higher frequencies than 18.9 GHz, a sharp HTM and relatively broad
LTM are split well and easily identified. } \label{raw09-27Tswp}
\end{figure}

\begin{figure}
\includegraphics[width=\linewidth]{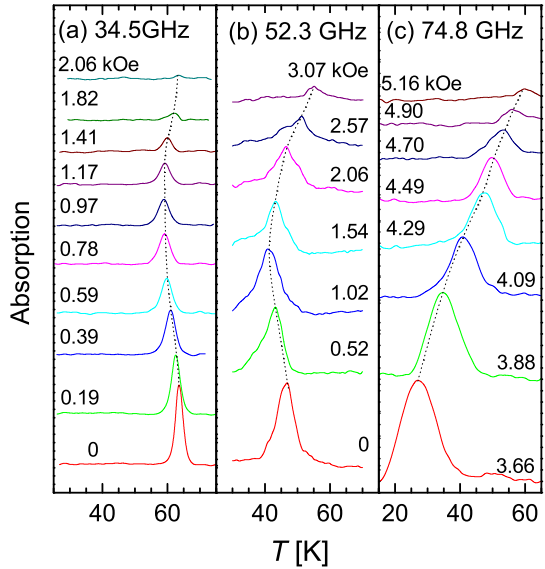}
\caption{
The resonance curves observed at 34.5 (a), 52.3 (b), and 74.8 (c)
GHz by sweeping temperature from above $T_c$. In this high
frequency region, only HTM was observed. The resonance drastically
shifts to lower temperatures and broadens as $\omega$ is
increased. At 74.8 GHz, HTM was obtained only above 3 kOe without
zero field resonance. } \label{raw34-75Tswp}
\end{figure}

Figure \ref{rawdata25}(a) shows microwave absorption curves at
25.5 GHz in various constant parallel magnetic fields
$H_{\parallel}$ obtained by TS measurements. Two clear resonance
lines with different characters are found at higher and lower
temperatures. As is seen clearly, there is a well-separated
temperature gap in-between. This feature strongly depends on the
microwave frequencies: the higher in frequency the larger
separation in temperature is observed. As will be shown later, the
lower temperature mode (LTM) quickly shifts towards the low
temperature side and disappears with increasing microwave
frequency $\omega$. On the other hand, with lowering frequency
below 30 GHz the LTM quickly shifts up and merges with the higher
temperature mode (HTM), which is explained in Fig.
\ref{raw09-27Tswp}. In general, the HTM in this frequency region
appears only in a relatively low field, approximately below 3 kOe,
as long as $\omega$ is lower than 55 GHz, which is close to the
zero-field and zero-temperature plasma frequency $\omega_p(0)$
described later. When $\omega$ goes beyond 33 GHz, only one
resonance is observed as shown in Fig. \ref{raw34-75Tswp}. In the
frequency region between 10 and 30 GHz both resonances come into a
play, resulting in a complicated phenomenon as a function of
temperature as well as magnetic field.

\begin{figure}[tb]
\begin{center}
\includegraphics[width=0.9\linewidth]{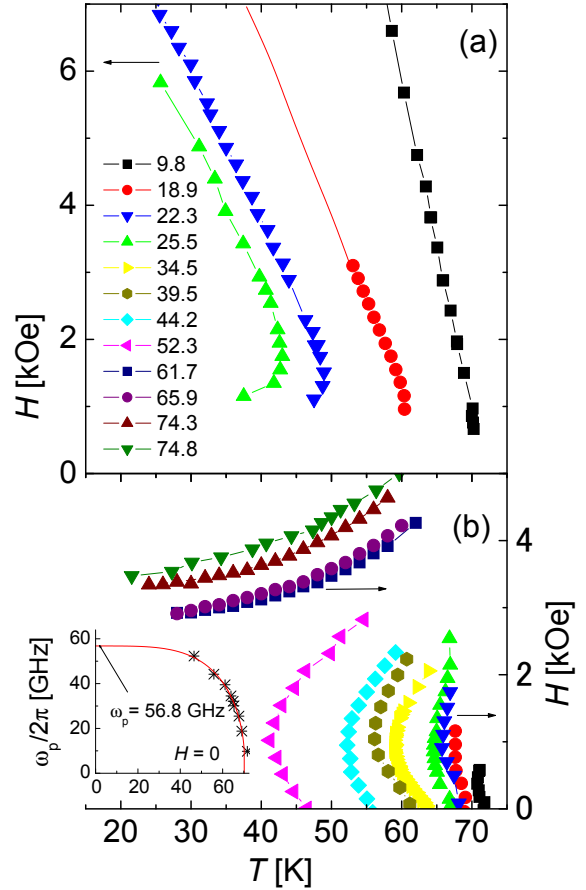}
\end{center}
\caption{
(Color online) (a): The resonance field-temperature plot for the
LTM resonance obtained at 9.8, 18.9, 22.3, and 25.5
 GHz.
(b): The resonance field-temperature plot for the HTM resonance
obtained at 9.8, 18.9, 22.3, 25.5, 34.5, 39.5, 44.2, 52.3, 61.7,
65.9, 74.3, and 74.8 GHz from right to left. The same symbols
represent data obtained at the same frequencies. Inset: The
$\omega_p$ vs. $T$ plot at zero field. Asterisks are experimental
data and the solid line is a curve fitted to the two fluid model.
} \label{H-T}
\end{figure}

\begin{figure}
\includegraphics[width=\linewidth]{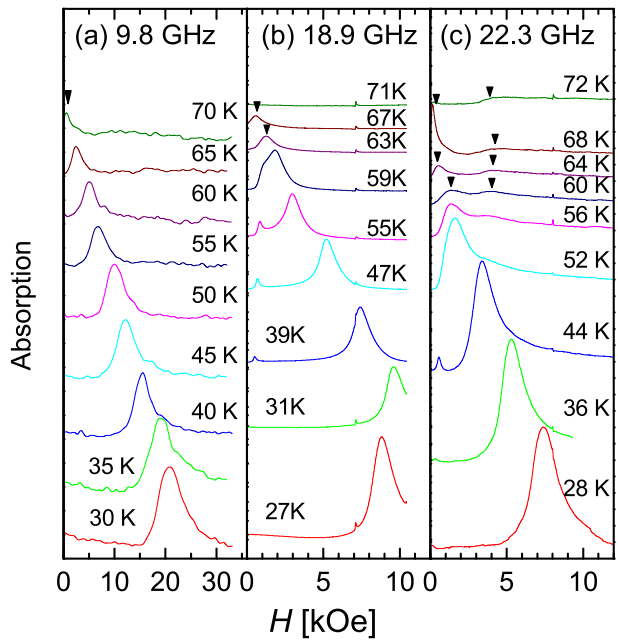}
\caption{
The resonance curves at 9.8 (a), 18.9 (b), and 22.3 (c) GHz by
sweeping magnetic field to $H=0$ after stabilizing temperatures at
the maximum fields. The resonance peaks pointed by inverted
triangles correspond to the HTM resonance.
 } \label{raw09-27Hswp}
\end{figure}

Focusing on the HTM, it behaves in a very unusual manner as seen
in Fig. \ref{rawdata25}(a), for example. In zero magnetic field a
sharp resonance with a symmetric lineshape with respect to
temperature was clearly observed. Since this resonance corresponds
to the zero-field mode argued previously, the temperature
dependence of the zero-filed plasma frequency $\omega_p(T)$ can be
described by the two-fluid model assuming the conventional Drude
model.~\cite{New3SC99}

In a finite parallel magnetic field below approximately 1 kOe, HTM
begins to shift slightly to lower temperatures. With further
increase of the magnetic field, however, it turns to shift
backward to higher temperatures at around 0.8 kOe as obviously
seen in Fig. \ref{rawdata25}(a) [this behavior is more clearly
displayed in Fig. \ref{H-T} (b), where the data obtained at many
frequencies measured are presented for the LTM]. The strong
absorption intensity of this HTM in zero field quickly decreases
as the field is applied. The linewidth in terms of temperature is
wider in finite fields than in the zero field.

It has been well-established that $\omega_p$ is always suppressed
by application of perpendicular magnetic
fields,~\cite{plasma2ka}\cite{Shi99a} because the Josephson
current due to coherence effect between layers is a decreasing
function of magnetic field.~\cite{Kos96} This requires that in
higher magnetic fields the resonance temperature must be lower in
order to gain the fraction of Josephson current which is reduced
by the perpendicular magnetic fields. It seems that this is not
the case in the Josephson plasma resonance in parallel fields to
the layers and this phenomenon certainly requires a new
explanation.

Such an unusual turnover behavior just mentioned above cannot be
explained either without making an unusual assumption that the
phase coherence between layers would increase further with further
increase of the parallel magnetic field approximately above 0.8
kOe or without introducing entirely new mechanisms for the
Josephson plasma resonance in the parallel magnetic field. It is
worth mentioning that this increasing tendency of the temperature
dependence of $H_\mathrm{res}$ toward higher temperatures seems
never to go beyond $T_c$ in zero field, although it goes beyond
the plasma resonance temperature $T_0$ in zero field at the
frequency $\omega$ as seen in Fig. \ref{H-T}(b). It fades away
quickly just below $T_c$. This surprising result may allow us to
speculate that the HTM is more stable in temperatures with the
formation of the Josephson lattice state than without Josephson
lattices. This unusual behavior has never been observed in the
Josephson plasma resonance in perpendicular magnetic fields
studied earlier and is certainly a new phenomenon only observable
in parallel magnetic fields.

In parallel magnetic fields, consecutive occupation of the
Josephson vortices in-between the CuO$_2$ layers is expected to be
realized along the $c$ axis as a function of increasing magnetic
field, preserving an isosceles triangular symmetry according to
the theoretical calculations.~\cite{Hu_00} This formation of
Josephson vortex lattices would cause the oscillatory coherent
Josephson current, which accordingly possesses the wave vector
$\bm{k}$ associated with the periodicity of the Josephson vortex,
and would provide the resonance at even higher temperatures in a
finite field than that in zero field according to the periodic
Josephson vortex arrangement.

At the intermediate field region between 1.3 and 2.2 kOe both the
HTM and LTM can be observed at the same magnetic field as seen in
Fig. \ref{rawdata25}(a). For instance, the data at 1.69 kOe
indicate the existence of both the HTM and LTM with 25.5 GHz at
(1.69 kOe, 66.2 K) and (1.69 kOe, 43.0 K), respectively. This
feature is observed only within a certain frequency region between
18 and 30 GHz, and only one resonance line can be found at the
other frequencies. This observation of two resonance modes
indicates a peculiarity in the the Josephson plasma resonance in
$\bm{H} \parallel ab$ unlike the one for $\bm{H} \parallel c$,
where only one resonance line as far as the longitudinal mode is
concerned, was observed to monotonically decrease with both
increasing temperature and magnetic field.


The resonance curves obtained by FS measurements are displayed in
Fig. \ref{rawdata25} (b). As the temperature is just below $T_c$
(= 70.2 K) a sharp resonance begins to appear ($T_0 \simeq$ 68 K)
from the zero field limit. The resonance rapidly grows with
decreasing temperature. As a few degrees Kelvin below $T_0$, two
resonance line can be distinguished: one at a lower field is
stronger in intensity than the other one which lies at a higher
fields and at the foot of the resonance of the stronger line
(pointed by arrows). This two-resonance feature can only be found
in 62 $\lesssim T \lesssim$ 66 K, below which the resonance
suddenly changes its character with broadening and with highly the
hysteretic nature as seen in Fig. \ref{rawdata25}(b). The
temperature range is wider at higher frequencies. For example, it
extends to 14 K at 52.3 GHz, above which frequency the two
resonance feature can no longer be observed.

As seen in Fig. \ref{rawdata25}(b) the absorption curves show a
considerable hysteretic behavior for the field being swept up and
down, especially below 45 K. This hysteretic feature is not
appreciable at high temperatures above 60 K whereas it is most
prominent in the intermediate temperature region between 10 and 45
K. It is noticed that there are several small resonancelike peaks
at the lower-field side in addition to the main large resonance.
However, this anomaly as well as even in the main peaks depends
strongly on the field sweeping direction and the sweeping rate.
Furthermore, it must be mentioned that some faint peaks are
observed in-between the HTM and LTM in the FS mode [at 52 K in
Fig. \ref{rawdata25} (b), for example], where there is no
resonance in the TS mode because of the gap between the HTM and
LTM as seen in Fig. \ref{rawdata25} (a). The peaks are, perhaps,
non-resonant absorptions due to the non-equilibrium effect of the
Josephson vortex state because the peak fields are almost equal to
the fields where the LTM is observed at the highest temperatures.
Therefore, we do not consider these faint peaks in the FS
measurement hereafter.

\subsection{Frequency Dependence
\label{sec:Freq}}
\subsubsection{Low-frequency region ($\omega/2\pi <$ 30 GHz)}

The typical resonance curves at 9.8, 18.9, and 22.3 GHz obtained
by TS measurements are shown in Figs. \ref{raw09-27Tswp}(a),
\ref{raw09-27Tswp}(b), and \ref{raw09-27Tswp}(c), respectively. At
the lowest frequency of 9.8 GHz, only one resonance line is
observed. The resonance line shape changes from asymmetric to
symmetric with increasing field below 0.5 kOe, indicating a
precursor effect of splitting into the HTM and LTM already at this
frequency region. In sharp contrast to this, the resonance
splitting is clearly observed at 18.9 GHz as shown in Fig.
\ref{raw09-27Tswp}(b), where the LTM resonance and the HTM
resonance appear below and above 64 K, respectively. The HTM
resonance becomes weak in intensity and is quickly smeared with
increasing field, and finally disappears at 1.0 kOe, while the LTM
resonance appears from the low-temperature side above about 0.7
kOe, then shifts to the higher-temperature side, and then turns
back to lower temperatures with increasing field with little
change in the lineshape up to 3 kOe. The larger separation between
the LTM and HTM resonances at higher frequencies indicates that
the LTM has a different frequency dependence from the HTM. The
integrated $H-T$ diagrams for most frequencies for both the LTM
and HTM are shown in Fig. \ref{H-T} by plotting the resonance
temperatures as a function of field.

The resonance curves obtained in FS measurements for various
temperatures at the same frequencies shown in Fig.
\ref{raw09-27Tswp} are shown in Fig. \ref{raw09-27Hswp}. All
resonance peaks displayed here belong the LTM except for the ones
indicated by the inverted triangles. The resonance peaks of the
LTM in increasing field lie at a lower field than the ones in
decreasing field, and the hysteresis observed in increasing and
decreasing magnetic fields becomes more significant at lower
temperatures and in lower fields. In comparison with the TS data,
the resonance peaks obtained in the case of decreasing field
coincide with the peaks in the TS measurements rather well, in
which more homogeneous Josephson vortex state is expected to
occur.

It is very unlikely that a complete lock-in vortex state is
realized without any pancake vortices in the actual experimental
condition, because the calculated angle to be allowed to form the
lock-in state by taking into account of the sample geometry (the
interlayer distance of 12 \AA and the length of the sample of 0.8
mm) would be of the order of $10^{-4}$ deg or less, which is much
smaller than the experimental angular resolution. Furthermore,
from the x-ray parallel-beam double-crystal method, it has been
known that our sample used for the present experiment has a width
of the rocking curve of the order of $0.025^\circ$. This means
that the crystallographic imperfectness of our sample, especially
the parallelism of the CuO$_2$ layers, gives an actual limitation
to the measurements. Therefore, we think that the Josephson vortex
state is always accompanied by some amount of the remaining
pancake vortices. In such a situation the vortex crossing lattice
state would be formed as first suggested by
Koshelev,~\cite{Kos99b} where the interaction between Josephson
vortices and pancake vortices may be attractive. Since the pancake
vortices are pinned strongly at low temperatures as seen in the
large hysteresis in magnetization, the Josephson vortex (JV)
system would be compelled to be pinned by the pancake vortices
through an attractive interaction. At the present stage of
knowledge we do not know to what extent this pinning effect is
important in the Josephson plasma phenomena except for the
hysteretic effect while the magnetic field is swept.

However, it is important to note that the complete JV state can be
established in a smaller sample with the dimensions less than $10
\mu\mathrm{m} \times 10 \mu\mathrm{m}$ in the $ab$ plane. This is
evidenced by the angular dependence measurement of the JV flow
resistance which shows a sudden rise at an angle within one degree
or so from the $ab$ plane, depending strongly on the sample size
and the intensity of the applied magnetic field.~\cite{LI-1}

\subsubsection{High frequency region}

Figures \ref{raw34-75Tswp}(a), \ref{raw34-75Tswp}(b), and
\ref{raw34-75Tswp}(c) depict resonance absorption curves at 34.5,
52.3, and 74.8 GHz, respectively, by TS measurements at various
fixed magnetic fields. In this higher-frequency region, only the
HTM is observed. As seen from Fig. \ref{raw34-75Tswp}, it is
indispensable to note that the HTM at 34.5 and 52.3 GHz has zero
field resonance, while it does not have zero field resonance at
74.8 GHz. This is because the HTM in the low-field region quickly
shifts to the low temperature side and disappears above
approximately 57 GHz. This can be seen more clearly in Fig.
\ref{H-T}(b), where the HTM is shown in the $H-T$ plot. This
zero-field resonance occurs only at a particular temperature $T_0$
at the corresponding $\omega$. The $\omega$ dependence of $T_0$
has been reported previously\cite{New3SC99,Gai99} and is explained
by using a simple two-fluid model with a temperature-independent
scattering rate.\cite{New3SC99} By extrapolating this temperature
dependence of the plasma frequency $\omega_p(T)$ to $T=0$ as shown
in the inset of Fig. \ref{H-T}(b), the inherent (zero temperature)
plasma frequency $\omega_p(0)/2\pi$ is estimated to be 56.8 GHz,
which yields $\lambda_c=217 \mu$m with $\epsilon =15$. This value
of $\lambda_c$ is quite reasonable for underdoped BSCCO with $T_c$
of 70.2 K and is used to quantify the field dependence of the HTM
in the later sections.

In the frequency range below but near $\omega_p(0)$, the dramatic
turnover behavior of the resonance temperature is observed as seen
in Fig. \ref{H-T}(b). The Josephson plasma resonance at low fields
below 1 kOe shifts to the lower temperature side and then turns
back to higher temperatures above 1 kOe as the magnetic field is
increased. A typical example is shown in the $H-T$ plot in the
left panel of Fig. \ref{H-TtoOmega-H} in the case of $\omega/2\pi
= 44.2$ GHz, where $T_0 = 55.7$ K and the shift in temperature
amounts to about 3.5 K below $T_0$ at about 1 kOe, then the
resonance peak sharply turns over toward the high-temperature
side, and reaches to 59 K, which is even more than 3 K higher than
$T_0$. The intensity of the resonance becomes weaker and weaker
with increasing magnetic field, and the resonance finally fades
away. As the frequency is increased, the whole curves including
$T_0$ moves towards low temperature as shown in the $H-T$ plot in
Fig. \ref{H-T}(b). However, it is interesting to note that the
disappearing temperature of the resonance after the turnover lies
in the temperature region between 55 and 65 K even above the zero-
field resonance temperature $T_0$. This surprising behavior is
very different from our understanding of the Josephson plasma
resonance for $\bm{H} \parallel c$, where the plasma frequency
{\it should} decrease with application of a magnetic field and
with increasing temperature as a result of the suppression of the
interlayer coherence $\langle \cos \varphi_{l,l+1} (H,T) \rangle$
due to the fluctuation of pancake vortices as formulated in Eq.
(\ref{JPR}).

At frequencies above $\omega_p(0)$, there is no longer the
resonance line at $H=0$ but only one resonance in a finite field
in the HTM. As shown in Fig. \ref{raw34-75Tswp}(c), the resonance
line can be observed only above about 3 kOe, and the resonance
occurs at higher temperatures in higher magnetic field until
reaching the limiting temperature of about 60 K around 5 kOe. This
means that the excitation energy of the HTM becomes higher at
higher temperatures. Since the superconducting phase coherence
should become weaker and weaker at higher temperatures, this
increasing excitation energy of the Josephson plasma in the HTM
needs a new expression if it is enhanced further by overcoming the
reducing temperature effect. As shown in Fig. \ref{H-T}(b), three
more data sets obtained at different frequencies are included.

\subsubsection{Overall features
\label{sec:overall}}

\begin{figure}[tb]
\begin{center}
\includegraphics[width=0.9\linewidth]{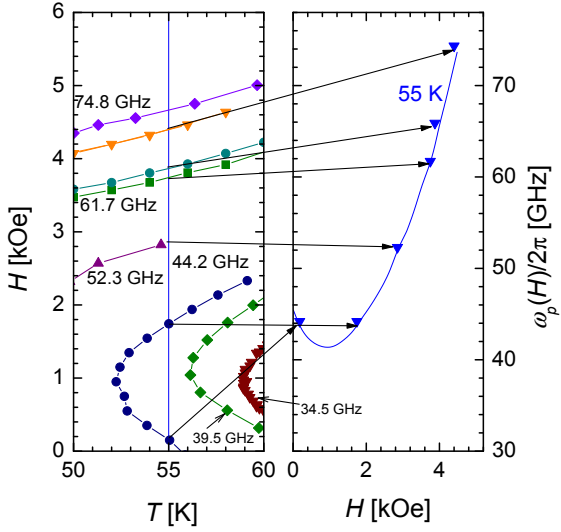}
\end{center}
\caption{
The reploted $H_\mathrm{res}$-$T$ diagram from the resonance
field-temperature plot shown in Fig. \ref{H-T} to the $\omega$-$H$
diagram at various frequencies at 55 K as an example.  }
\label{H-TtoOmega-H}
\end{figure}

\begin{figure}
\includegraphics[width=\linewidth]{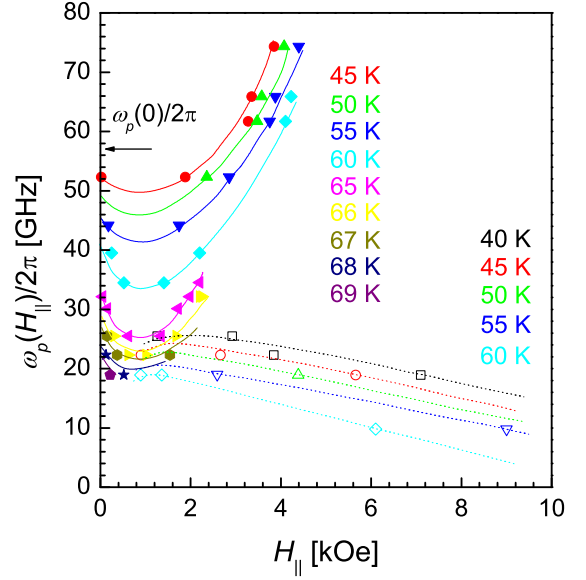}
\caption{
(Color online) The frequency-field diagram for HTM and LTM. Solid
symbols denote HTM at 45, 50, 55, 60, 65, 66, 67, 68 and 69 K, and
open symbols denote LTM at 40, 45, 50, 55 and 60K from top to
bottom. Solid and dotted curves are guide for the eyes.
\label{omega0-H} }
\end{figure}

\begin{figure}
\includegraphics[width=\linewidth]{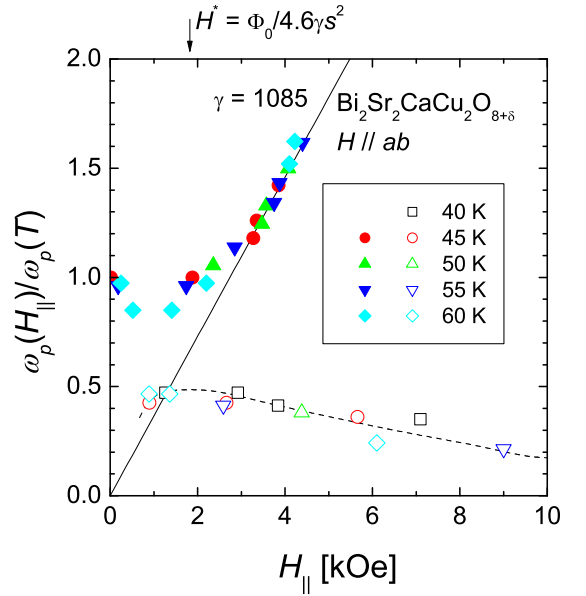}
\caption{
(Color online) $\omega_p(H_{\parallel})/\omega_p(T)$ of the both
modes at 40, 45, 50, 55 and 60 K are shown. Solid and open symbols
 denote HTM and LTM, respectively, and a solid line is given by
Eq. (\ref{ParallelPlasma}) with $\gamma=1085$. \label{omega1-H} }
\end{figure}

In order to construct the $\omega_p-H$ diagram, which is a direct
indication of the $H_{\parallel}$ effect on the excitation modes,
the following procedure is applied: Since the most of experiments
have been done in the TS mode at a fixed magnetic field at a
certain $\omega$ and the $H-T$ diagram was constructed as shown in
Fig. \ref{H-T}, one should convert this to the $\omega_p-H$ plot
by taking the resonance points from corresponding $\omega$ at
various temperatures as shown in Fig. \ref{H-TtoOmega-H}. The
resonance frequency is plotted as a function of magnetic field for
nine different temperatures in Fig. \ref{omega0-H}. It is noted
that the plots in Fig. \ref{omega0-H} were extracted from
resonance field versus temperature plots at all measured
frequencies (partly not shown in Fig. \ref{H-T}) .

It is clear that there are two different type of resonance modes
with very different field dependent characteristic features: one
corresponds to the HTM, which starts with $\omega_p(0)$ at zero
field; then, the frequency grows monotonically and almost linearly
after passing a shallow minimum in the $\omega_p$ vs $H$ diagram
as the field is increased. On the contrary to this, the LTM shows
a weak initial increasing behavior with increasing magnetic fields
and then turns to the gradual decreasing behavior after passing a
broad maximum. This LTM seems to be disappear at the zero-field
limit.

As mentioned above, the HTM deviates from linear behavior and
tends to have a minimum at 0.5 -- 1.0 kOe, and then merges with
the zero field Josephson plasma mode. Since this HTM is directly
connected to the zero-field Josephson plasma mode, the HTM is
confirmed to be the Josephson plasma mode in the Josephson vortex
state in a parallel external magnetic field. It is noted that the
initial decrease of $\omega_p(H_{\parallel})$ is stronger at
higher temperatures. On the other hand, the LTM exhibits a broad
peak at around 1.5 kOe and then tends to decrease again with
decreasing field. It is important to note that this LTM disappears
at the limit of zero magnetic field. This experimental fact
implies strongly that the existence of JV's is crucial to the LTM.

In Fig. \ref{omega1-H}, $\omega_p(H_{\parallel})$ normalized by
the zero field plasma frequency at a given temperature derived
from the inset of Fig. \ref{H-T}(b) is shown for five typical
temperatures in Fig. \ref{omega0-H}. As seen in Fig.
\ref{omega1-H} it is interesting to point out that both the HTM
and LTM are scaled well to two single curves by using
$\omega_p(T)$.

\subsection{Universality of Two Excitation Modes
\label{sec:Doping}}

\begin{figure}
\includegraphics[width=\linewidth]{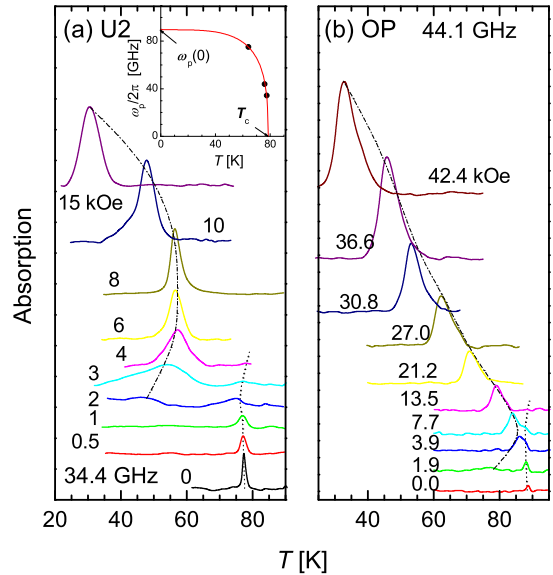}
\caption{
The resonance data obtained (a) at 34.6 GHz in U2 ($T_c$ =76.8 K)
and (b) at 44.1 GHz in OP ($T_c$ =91.0 K). Two resonances and the
temperature gap are found in both samples. Separation of two modes
at 4 kOe in U2 (a) is similar as the data at 22.3 GHz and 1.6 kOe
in U1 (Fig. \ref{raw09-27Tswp} (c)). The inset of (a) shows
temperature dependence of the zero-field plasma frequency in U2.
Dotted lines trace the HTM resonance and is guide for the eyes,
which shows turnover behavior as well as HF mode in U1. }
\label{rawU2_35-OP_44}
\end{figure}

The microwave measurements have been performed in a similar manner
in other crystals with higher doping levels (U2 and OP samples).
The essential features described above were unchanged: the
characteristic two modes HTM and LTM exist with qualitatively
similar temperature as well as magnetic field dependencies. A
typical example of the experimental data are shown in Figs.
\ref{rawU2_35-OP_44}(a) and \ref{rawU2_35-OP_44}(b) for an
underdoped U2 sample and for an optimally doped OP sample,
respectively. Comparing these data with the data shown in Figs.
\ref{raw09-27Tswp} and \ref{raw34-75Tswp}, one can easily find
that the separation between the HTM and LTM for U2 is smaller than
that of U1 at a similar frequency. For example, the data in Fig.
\ref{rawU2_35-OP_44}(a) (at 34.4 GHz) can be compared with the
data in Fig. \ref{raw34-75Tswp}(a) (at 34.5 GHz) and are rather
similar to the those obtained at 22.3 GHz shown in Fig.
\ref{raw09-27Tswp}(c). Assuming that these two data sets obtained
at 22.3 and 34.5 GHz provide the same ratio of
$\omega/\omega_p(0)$ to each sample, $\omega_p(0)/2\pi$ for U2 can
be derived as 88.1 GHz. This estimated value of $\omega_p(0)$ is
consistent and agrees rather well with the value extrapolated by
the temperature dependence of the zero-field Josephson plasma
frequency $\omega_p(T)$ as shown in the inset of Fig.
\ref{rawU2_35-OP_44}(a), although the number of experimental
points are only a few.

In the case of the OP sample, typical experimental data are shown
in Fig. \ref{rawU2_35-OP_44}(b) at 44.1 GHz, where no clear
splitting of HTM and LTM is observed. Applying the similar scaling
in this case (for 18.9 GHz in U1), a value of $\omega_p(0)/2\pi
\approx 130$ GHz is derived. This value is close to the one
obtained for an optimally doped Bi2212 single
crystal.~\cite{Gai99} Thus we suggest that the resonance splitting
can be noticeably observed at $\omega \gtrsim \omega_p(0)/3$ and
the magnitude of the $T$-gap is larger for $\omega/\omega_p(0)$
being closer to unity. In contrast, $\omega$ is much smaller than
$\omega_p(0)$, and the LTM merges to HTM as observed at 9.8 GHz in
U1, resulting in a single branch of the $H-T$ plot.


In previous studies, Matsuda {\it et al}.~\cite{Mat97a} and Tsui
{\it et al}.~\cite{Tsu97} have reported that the Josephson plasma
resonance field has a sharp decrease near $\theta=0$ ($\bm{H}
\parallel ab$). In particular, Matsuda {\it et al}. found the
disappearance of resonance in an underdoped crystal ($T_c=87$ K)
through measurements of the angular dependence of the FS data and
argued the existence of a collective mode of the {\it Josephson
vortex lattice} which lies above the experimental microwave
frequency ($\omega/2\pi=45$ GHz) at zero field and increases with
parallel magnetic fields on the basis of a theory by Bulaevskii
{\it et al.}~\cite{Bul97}. Judging from our results, this
vanishing of the resonance observed by Matsuda {\it et al.} can be
interpreted differently because their sample is underdoped crystal
and the measurement was done at 36 K. Their experimental
conditions seem to be in the gap region between the HTM and LTM.
Furthermore, according to their results the disappearance was not
observed in an optimally doped crystal ($T_c=89.5$ K). This also
seems to be consistent with the systematics which we found here
although they attributed it to the misalignment of the magnetic
field. From an experimental point of view, no matter what the
theoretical interpretation may be, the observed dramatic change of
the Josephson plasma resonance near $\theta = 0$ is strongly
suggestive for the excitation of the new phase collective modes in
intrinsic Josephson junctions.


\section{Origin of excitation modes\label{Origin}}

\subsection{High Temperature Mode}

\begin{figure}
\begin{center}
\includegraphics[width=\linewidth]{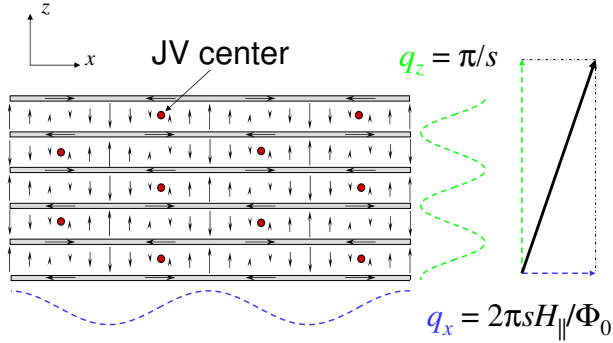}
\end{center}
\caption{
A schematic picture of propagation of the tilted plasma wave. The
interlayer phase difference along the $ab$ plane and the $c$ axis
are spatially modified as shown by dotted and broken waves,
respectively. The propagation vector of the plasma wave comprised
of the transverse component $q_x$ and the longitudinal component
$q_z$ is depicted as a thick arrow. \label{JV+plasma} }
\end{figure}

As discussed by Fetter and Stephen in single
junctions,~\cite{Fet68} propagating Josephson plasma waves are
strongly modified by the JV's, resulting in the Josephson plasma
mode having a linear field dependence according to the linear
increase of the reciprocal lattice vector of the JV array,
$k_H=2\pi s H_{\parallel}/\Phi_0$, where $\Phi_0$ is the flux
quantum. This indicates that a periodic arrangement of JVs is
crucial to have the propagating Josephson plasma mode. We first
assume that JV's penetrate all block layers and form isosceles
triangular lattice with the lattice constants along the $c$ axis
and the $ab$ plane being $2s$ and $\Phi_0/sH_{\parallel}$ in
high-$H_{\parallel}$ region, respectively, as shown in Fig.
\ref{JV+plasma}, although for layered superconductors the argument
has not been completely settled yet. A theoretical calculation by
Ichioka~\cite{Ich95} predicts that such a dense JV lattice can be
realized above $H^*=\sqrt{3}\Phi_0/8\gamma s^2$, which gives 1.8
kOe for $\gamma=1085$ obtained from $\lambda_c=217 \mu$m and
$\lambda_{ab}=2000$ \AA. This value is rather in good agreement
with the magnetic field, where the HTM begins to tend to the
linear relation and the LTM has a broad maximum. Hence, it is
reasonable to think that the HTM above $H^*$ may be the Josephson
plasma resonance mode in the dense JV lattice.

Considering a Josephson plasma wave traveling in the isosceles
triangular JV lattice, the $\bm{k}$ vector of the Josephson plasma
mode corresponds to the primitive reciprocal lattice vector of the
JV lattice $\bm{q}$, which consists of the constant $c$-axis
component $q_z=\pi/s$ and the $H_{\parallel}$ proportional
$ab$-plane component $q_x=2\pi s H_{\parallel}/\Phi_0$, as
depicted in Fig. \ref{JV+plasma}. Here, $y$ and $z$ axes are taken
parallel to the JV and $c$ axis, respectively, and $x$ is taken
perpendicular to the $y$ and $z$ axes. Thus, Josephson plasma
waves can be excited at all possible wave vectors as of the
mixture of the longitudinal plasma with $k_z=q_z$ and the
transverse plasma with $k_x=q_x$. The plasma frequency is
determined by the dispersion relation along $\bm{q}$ lying between
the longitudinal and transverse plasma modes and increases with
magnetic field because of the linear increase in $q_x$. It should
be noted that the dispersion is very close to the longitudinal
dispersion for $q_x \ll q_z$.

Bulaevskii {\it et al}. have proposed a formulation of the plasma
resonance in a layered superconductor on the basis of the single-
junction model,~\cite{Bul97} and Koshelev and Machida have derived
excitation spectra by considering the longitudinal coupling
effect.~\cite{Kos01,Mac01} In the high field limit, where all
block layers are occupied by JV's forming an isosceles triangular
lattice, the peak frequency of the dissipation spectrum,
corresponding to the resonance peak, as a function of parallel
magnetic field is formulated as
\begin{equation}
\frac{\omega_p(H_{\parallel})}{\omega_p(0) }= \frac{\pi \gamma
s^2}{\Phi_0}H_{\parallel} . \label{ParallelPlasma}
\end{equation}
Using $\gamma=1085$, this equation provides an excellent agreement
with the experimental data above 3 kOe at all temperatures, as
shown in Fig. \ref{omega1-H}. It is also consistent that the
linear field dependence of the HTM is violated below $H^*$. We
therefore conclude that the linear increase in
$\omega_p(H_{\parallel})$ of the HTM is due to an increase in
$q_{ab}$ of the JV lattice, while $q_c$ is fixed on $\pi/s$
because of the intrinsic pinning. This result suggests that the
excited plasma frequency can be easily controlled by adjusting the
magnetic fields parallel to the layers even above $\omega_p$.

\subsection{Low Temperature Mode}

The LTM lies at $\omega \lesssim \omega_p/2$ and slightly
decreases with $H_{\parallel}$ as displayed in Fig.
\ref{omega1-H}. The LTM in crystals with higher doping tends to
appear at higher frequencies although the whole $\omega_p-H$
diagram has not been established yet. In under doped YBCO, a
similar excitation mode was observed around $\omega_p/2$ with
slight negative $H_{\parallel}$ dependence and is referred to the
$\alpha$ mode~\cite{Koj02}. The $\alpha$ mode cannot be observed
after zero-field cooling, suggesting the considerable influence of
pinning of the JV's on the collective oscillation mode. This may
also be the case in Bi2212 as shown in Fig. \ref{rawdata25}(b) and
as well as in Fig. \ref{raw09-27Hswp}, where a large hysteretic
effect is observed. Assuming that our LTM corresponds to their
$\alpha$ mode, the low-lying collective mode should strongly
depend on $\omega_p$, which is a material parameter and given by
the strength of the Josephson coupling between CuO$_2$ bilayers.

It is well known that the lowest collective mode of a Josephson
vortex system is the vortex sliding mode with
$k_x=0$.~\cite{Fet68} The sliding mode becomes gapless in an ideal
system without pinning and couples with a dc homogenous electric
field perpendicular to the superconducting layers. Since the
pinning of Josephson vortices cannot completely be removed in
actual samples, the lowest mode should have a nonzero frequency.
In a junction with a periodic pinning for instance, the Josephson
critical current depends periodically on $x$.~\cite{Bul97}

Since the hysteresis obtained by the FS measurements indicates
that the JV pinning strongly affects to the LTM, we once discussed
the {\it pinned vortex sliding} scenario as a possible origin of
the LTM.~\cite{ISS01} However, our finding that LTM tends to
become higher in more highly doped samples with higher $\omega_p$
cannot be explained by this scenario, because it is not natural
that the collective mode hardened by the pinning effect can be
scaled by $\omega_p$. Therefore, the vortex sliding model is not
suited for and not likely for the origin of LTM in the framework
of conventional knowledge, and needs for a new explanation.

In order to reveal the origin of the LTM, it is required to
introduce an intrinsic stacking effect of intrinsic Josephson
junctions because previous theoretical models giving the vortex
sliding mode are based on the single junction model, where the
correlation of the JV lattice along the $z$ axis is neglected. The
vortex sliding mode in layered superconductors which we have in
mind means coherent vortex motion parallel and perpendicular to
the layers with $k_x=k_z=0$. Here, we may consider a collective
mode with $k_x=0$ but $k_z \neq 0$ at a finite frequency.
Considering an antiphase oscillation of JV arrays between adjacent
junctions with $k_z=\pi/s$ and $k_x=0$, for instance, the
oscillation is a sharing motion between adjacent JV arrays as a
result of dynamic phase oscillations not only between adjacent
layers but also next-adjacent layers, so that the frequency of the
oscillation can be considered to be scaled by $\omega_p$ because
the interaction between JV's in adjacent junctions is to be
governed by the Josephson coupling as the first approximation.

Very recently, Koyama reported theoretical calculations on the
Josephson plasma resonance in the JV lattice.~\cite{Koy03} After
numerical calculations on analytically derived equations, a
collective mode which lies about $\omega_p(0)/2$ in the low-field
limit and slightly decreases with increasing field was obtained.
He argues that the collective mode is considered as an anti-phase
oscillation mode of JV's which originates from the strong charge
coupling between junctions, and it is also responsible for the
longitudinal Josephson plasma mode. This picture seems to be most
appropriate as a candidate for the origin of the LTM although no
direct experimental proof has been obtained.

For deeper understandings of the LTM, further theoretical and
experimental investigations are needed in the low-frequency
region. We think that an investigation of the JV flow branch in
the current-voltage characteristics along the $c$ axis is a
complementary experiment to elucidate the JV behavior in the
layered superconductors in the low-frequency region.

\section{Conclusions}
We have clearly identified {\it two} microwave excitation modes
with a temperature-dependent gap in parallel magnetic fields in
single crystal Bi$_2$Sr$_2$CaCu$_2$O$_{8+\delta}$. It is shown
that the strong coupling effect between the Josephson plasma and
JV lattice is responsible for these two modes. The high frequency
mode is attributed to the Josephson plasma mode propagating along
the primitive reciprocal lattice vector of the JV lattice. Making
use of this interplay, we may be able to explore electromagnetic
waves in the frequency range between the superconducting gap and
$\omega_p$. The low frequency mode is attributed to the new
collective mode which has never been observed before and seems to
be unique for layered superconductors. It is considered that the
Josephson vortex collective oscillation mode with finite $k_z$
contributes to the phase oscillations, although a quantitative
full understanding has not been obtained.

\begin{acknowledgments}
We would like to thank Drs. M. Machida, A. E. Koshelev, T. Koyama,
H. Matsumoto, K. M. Kojima and N. F. Pedersen for their fruitful
discussions and critical comments. This work has been supported by
the Grant-in-Aid for Young Scientists (B), No. 14740201 at
FY2002-2003, and 21st Century Center of Excellence (COE) Program
at University of Tsukuba under MEXT, Japan.
\end{acknowledgments}

\bibliography{MyBibs,MyPubs}

\newpage

\end{document}